\documentclass[sigconf,natbib=true,anonymous=false]{acmart}

\AtBeginDocument{%
  }

\usepackage{multirow}
\usepackage{amsmath}
\usepackage{graphicx}
\usepackage{subcaption}
\usepackage{algorithm}
\usepackage{algpseudocode}
\usepackage{makecell}
\usepackage[colorinlistoftodos]{todonotes}

\acmConference[]{}{}{}

\setcopyright{none}
\renewcommand\footnotetextcopyrightpermission[1]{} 

\begin{document}
\title{Enhancing Healthcare Search Intent Recognition with Query Representation Learning and Session Context}

\author{Chen Lin}
\affiliation{%
  \institution{Emory University}
  \city{Atlanta}
  \state{Georgia}
  \country{USA}}
\email{clin366@outlook.com}

\author{Harshita Jagdish Sahijwani}
\affiliation{%
  \institution{Emory University}
  \city{Atlanta}
  \state{Georgia}
  \country{USA}}
\email{harshita.jagdish.sahijwani@emory.edu}

\author{Madhav Sigdel}
\authornote{Data curation, Writing - review \& editing, Supervision.}
\affiliation{%
  \institution{Kaiser Permanente}
  \city{Oakland}
  \state{California}
  \country{USA}}
\email{Madhav.X.Sigdel@kp.org}

\author{Song Aslan}
\authornote{Data curation, Writing - review \& editing.}
\affiliation{%
  \institution{Kaiser Permanente}
  \city{Oakland}
  \state{California}
  \country{USA}}
\email{Song.X.Aslan@kp.org}

\author{Priya Gopi Achuthan}
\authornote{Data curation, Writing - review \& editing.}
\affiliation{%
  \institution{Kaiser Permanente}
  \city{Oakland}
  \state{California}
  \country{USA}}
\email{Priya.X.Gopiachuthan@kp.org}

\author{Monica D. Skidmore}
\authornote{Supervision.}
\affiliation{%
  \institution{Kaiser Permanente}
  \city{Oakland}
  \state{California}
  \country{USA}}
\email{Monica.D.Skidmore@kp.org}

\author{Eugene Agichtein}
\affiliation{%
  \institution{Emory University}
  \city{Atlanta}
  \state{Georgia}
  \country{USA}}
\email{eugene.agichtein@emory.edu}

\renewcommand{\shortauthors}{Sahijwani et al.}


\begin{abstract}
Classifying the intent behind healthcare search queries is crucial for enhancing online healthcare information delivery. The intricate nature of medical search queries, coupled with the limited availability of high-quality labeled data, presents significant challenges for developing efficient classification models. Previous studies have exploited user interaction data, such as user clicks from search logs, and have employed pairwise loss functions to model co-click behavior for query representation learning. However, many health queries can have multiple intents, leading to ambiguous or divergent click behavior. Furthermore, relying on the most popular intent inferred from global statistics based on aggregate user behavior can result in inconsistencies and reduced performance. This is particularly evident when classifying query intent within specific search sessions. To address these limitations, our work enhances query representation learning by clustering similar queries, thereby improving the ability to capture nuanced user intents. Additionally, we introduce a novel loss function designed to handle multi-intent scenarios, resulting in a more scalable and accurate learning procedure. Furthermore, we quantify the ambiguity of health queries and the misalignment between global search intents and those discerned from individual sessions by introducing the concordance rate (CR) score. We also demonstrate a simple and effective method for incorporating our learned query representation into contextual, session-based search intent classification. Our extensive experimental results and analysis on two real-world search log datasets, namely a Health Search (HS) dataset from Company A and the publicly available TripClick dataset, demonstrate that our approach improves intrinsic clustering metrics for query representation learning. It also enhances accuracy for subsequent search intent classification tasks.
\end{abstract}




\maketitle

\section{Introduction}

Health domain search engines rely heavily on discerning user intent to provide relevant results \cite{jansen2007determining, wang2022recognizing}. Users often input a variety of search queries such as symptoms, drugs, specific doctors, or health insurance information, expecting the search engine to comprehend these queries and accurately provide users with documents or web pages matching their needs. The categorization of user search intent (i.e., search intent classification) enables search engines to provide organized and relevant results, which further leads to higher user satisfaction.  


\begin{table}[h]
\centering
\footnotesize
\caption{Examples of medical search queries with corresponding search intents.}
\vspace{-0.3cm}
\label{tab:medical_search_queries}
\begin{tabular}{ll}
\toprule
\textbf{Medical Search Query} & \textbf{Search Intent} \\
\midrule

lice treatment & Seeking drug and wellness info \\
\hline
bd nano 2nd gen pen needle 32 gauge x 5/32 & Seeking drug information \\
\hline
hawaii advance directive form & \makecell[l]{Managing health accounts;\\ seeking wellness information} \\
\bottomrule
\end{tabular}
\end{table}

Detecting user search intent in general, and in particular in the health domain, is a challenging problem and has been an active area of research for many years. 
The main reason for the difficulty in predicting user intent is that medical search queries are inherently ambiguous due to conflation of specialized and colloquial terms, and can be difficult to interpret without context even by human annotators
 \cite{wang2015query, wang2017combining}, as shown in Table \ref{tab:medical_search_queries}. 

To address this problem, we build upon and expand on the general approach of representation learning. Specifically, we observe that the search logs in the search engine contain substantial information in user search intents \cite{agichtein2006improving, white2010predicting, bennett2012modeling}. For example, the co-click queries could be used as weak supervision for queries sharing similar search intent and the clicked document or web page annotations could also indicate users' search needs. 

Hence, a well established strategy for query intent modeling using search logs is to harness implicit user feedback based on user click behavior for learning query representations \cite{zhang2019generic}. This approach usually involves constructing the query pairs that lead to the same click (termed "co-click" queries) as indicative of similar user intent. Prior studies applied contrastive learning for utilizing the co-click query pairs as weak supervision, where they use a pairwise loss function that ensures that positive (co-click) query pairs are closer to each other, while negative query pairs are farther in their representation space. We review prior work in this area in Section~\ref{sec:related_work}. 

However, the effectiveness of this overall approach can degrade in the presence of ambiguous queries (i.e., those for which multiple intents are possible), and for situations like Health where a specialized health search engine may receive only a fraction of the click volume of general-purpose search engines. 

Furthermore, the approach of learning single query embeddings for single-label intent recognition may not suffice to represent the multifaceted nature of user queries \cite{yuan2023multi}. As shown in Table \ref{tab:medical_search_queries}, in the case of a user searching "lice treatment", the intent might span multiple categories, such as seeking drug information ("drug info") and treatment methods ("health wellness"), underscoring the need for a more comprehensive recognition strategy. Given the limitations of single-label models in capturing the full spectrum of user intents, a shift to multi-label intent recognition becomes essential.

Therefore, a natural question is, can we make the representation learning more robust and effective for the multi-faceted nature of Health queries, and the inherent noise in click logs from small-scale search engines? Stated differently, we aim to investigate: \textbf{RQ 1: How to improve multi-faceted query representation for Health search, by learning from moderately sized click logs?} 

To address RQ1, we introduce a novel multiset loss function, specifically designed to address the inherent ambiguity in user click logs for more accurate user intent learning. Unlike traditional methods, our approach significantly enhances the learning process by utilizing clicked document annotations as the weak supervision. By implementing a clustering-based learning approach, we can effectively harness the rich information embedded in user click logs, thereby allowing for a clustering-driven representation of search queries. This approach not only addresses the previously mentioned challenges but also directly facilitates the downstream search intent classification tasks.

While it is possible to effectively learn a {\em global} query intent representation, the {\em individual} users may have a different intent for a query within a specific search session. As shown in Table \ref{tab:query_misalignment}, there is a general disagreement between global intents from search logs and session-specific individual intents for identical search queries. This disparity underscores the complexity inherent in accurately understanding user intent based on isolated queries as opposed to considering the entire session context \cite{hashemi2020guided, ortiz2022exploring}. Such observations motivate the research into the refinement of search query representation in the context of user's search sessions. Therefore, we propose our second research question: \textbf{RQ2: How to enhance search query representation from search context for session-based search intent?}  

In summary, we propose a novel multiset loss function to leverage moderately sized search logs, especially the clicked document annotations, to learn effective query representations for Health search. The multiset loss function is based on the cosine similarity between the query embeddings and the centroid of the document set embeddings. To validate our approach, we apply it to two real datasets. The first is the Health Search dataset, sourced from Company A's internal search engine \footnote{Company details withheld for anonymity}. This search engine helps users find health wellness information as well as to meet specific healthcare needs like finding doctors and accessing records. The second dataset is the TripClick dataset \cite{rekabsaz2021tripclick}, a publicly available dataset of search engine log files. We evaluate the effectiveness of our approach on query embedding clustering and search intent classification. We further incorporate the session context into the query intent classification model to enhance the classification performance in individual search sessions. Our contributions are:
\renewcommand{\labelitemi}{$\bullet$}
\begin{itemize}
\item A novel multiset loss function to leverage search logs, especially the clicked document annotations, to learn the query intent representations for Health search.
\item Thorough experiments and analysis to demonstrate the effectiveness of our approach on two real-world datasets.
\item Evidence of the generality and robustness of our query representation approach, by demontrating that it can be used 
to predict session-level intent more accurately than the previous representations. 
\end{itemize} 
Through our extensive experiments and analysis, we demonstrate that our approach outperforms the prior state of the art, and can be easily replicated and adapted. While the motivating use cases and our experiments are focused on the Health search domain, the methods we introduce are general, and can be applicable to other settings where queries tend to have multiple interpretations, and a specialized search engine is beneficial (e.g., for privacy reasons). 


\begin{table*}[h]
\centering
\footnotesize
\caption{Disparity between global and session-level search intents for the same search queries}
\vspace{-0.3cm}
\begin{tabular}{p{3cm}p{3cm}p{3cm}p{3cm}p{3cm}}
\toprule
\textbf{Search Query (Step 1)} & \textbf{Search Query (Step 2)} & \textbf{Search Query (Step 3)} & \textbf{Global Intent (Step 3)} & \textbf{Session-Specific Intent (Step 3)} \\ 
\midrule
"healing" & "acupuncture" & "chiropractor" & Scheduling medical appointments; dealing with billing and coverage & Seeking wellness information \\
\hline
"online visit" & "e visit" &  "zofran" & Seeking drug information & Seeking communication information \\
\bottomrule
\end{tabular}
\label{tab:query_misalignment}
\end{table*}

\begin{figure}[htbp]
    \centering
    \includegraphics[width=\linewidth]{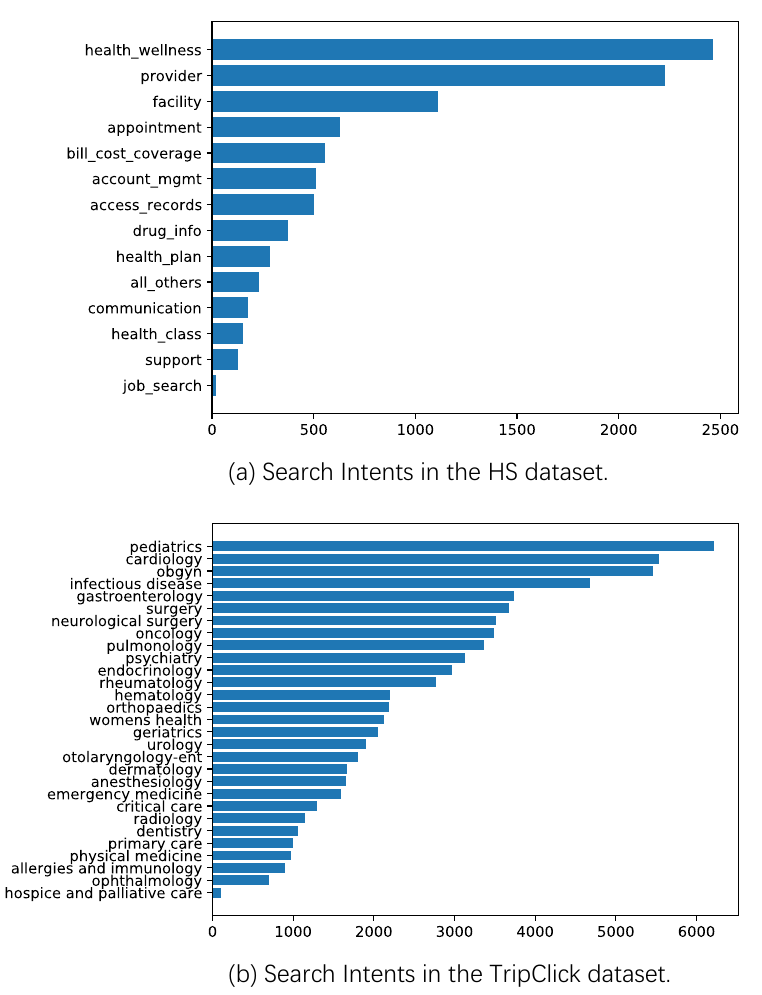}
    \vspace{-0.7cm}
  \caption{Comparative analysis of intent distributions in HS and TripClick datasets, highlighting distinct user search preference over different intents with specific intent types and frequencies.}
  \label{fig:label_distributions}
\end{figure}

\section{Related Work}
\label{sec:related_work}
In this section, we first present a short review on query intent understanding in section \ref{sec:query_intent_understanding}. Then, we briefly review query representation learning research in section \ref{sec:query_rep_learning}. 
\subsection{Query Intent Understanding} 
\label{sec:query_intent_understanding}
Query intent understanding is important for retrieving relevant results and has been widely studied \cite{broder2002taxonomy,shen2006building,broder2007robust,yin2010building,hashemi2021learning,yuan2023multi}. 
One of the earliest works on user intent analysis by \cite{broder2002taxonomy} classified user intents into three categories: navigational, informational, and transactional. 
Yin et al. in \cite{yin2010building} propose a method to learn a hierarchical taxonomy of the generic search intents for a class of name entities, relying solely on search logs. 

Since web search queries are short and ambiguous, contextual information is necessary for accurate intent recognition \cite{zamani2017situational}.
Broder et al. in \cite{broder2007robust} employ the pseudo relevance feedback paradigm and use the top search results to gain additional insights for query interpretation. The approach leverages document labels of the search results to infer the query intent. We follow a similar approach, but use the document classes of clicked documents instead of top k results to cluster queries by intent.

Previous studies have applied statistical learning and machine learning models to classify search queries into different intent categories \cite{Matt:2013, nalisnick2016improving,hu2009understanding, xiong2017end, rahman2013search}. Most recently, given the rapid development of natural language processing(NLP) models, more and more studies start to use pertained large language models (LLMs)  for query phrase representation and achieve promising results on this task \cite{srinivasan2022quill}. Contextual search enhances search results by considering search query's preceding context. Search logs have been suggested as one source for contextual information and many contextual search approaches have been proposed \cite{white2010predicting, bennett2012modeling, vuong2024predicting, yuan2023multi, hashemi2020guided, kong2015predicting, sen2021know}.

\subsection{Query Representation Learning}
\label{sec:query_rep_learning}
Recent studies have utilized the embeddings from the transformer architectures for representation learning to understand the context and improve the search relevance \cite{devlin2018BERT,lee2020bioBERT, cer2018universal, xiong2017end, zamani2017relevance}. Most recently, Zhang et al. in \cite{zhang2019generic} demonstrate that weak supervision through click data in combination with fine-tuning on human-labeled paraphrase classification task is an effective method of learning query representations. Wang et al. in \cite{wang2022recognizing} focus specifically on medical queries. To account for the noise and lack of specificity of search queries, they propose a query encoder that also encodes external semantic knowledge from a medical KG, syntactic knowledge from POS tags, and generic knowledge captured from large corpus. Through previous studies \cite{zhang2019generic, wang2022recognizing}, the pairwise loss optimization approach has shown promising results in learning the search query representations from the user click log data and improves the performance of the downstream search intent classification tasks. While contrastive learning with pairwise loss functions demonstrates its effectiveness across various domains, such as face recognition and clustering in computer vision \cite{schroff2015facenet}, its application on user clicks often results in the duplication of co-click pairs and noisy pairs due to random user clicks. 
To account for this, Zhang et al. in \cite{zhang2019generic} applies computationally efficient convolutional neural network (CNN) layers \cite{shen2014learning} and Bi-GRU layers to learn the query representations from the user click log data and fine-tune the model on the human-labeled paraphrasing task. However, it remains a challenge to learn the query representations from the user click logs where human-labeled data is limited, which is often the case for small-scale search engines.

\begin{figure*}[ht!]
\centering
\includegraphics[width=0.95\textwidth]{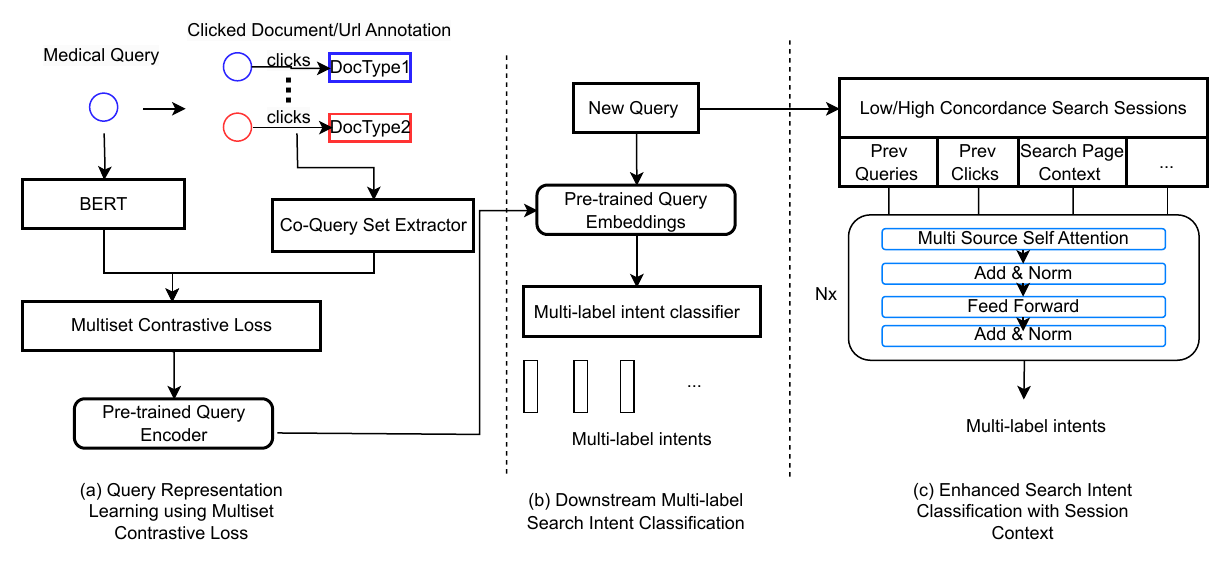}
\vspace{-0.5cm}
\caption{A comprehensive approach for query representation learning and intent classification: (a) illustrates the process of leveraging Multiset Contrastive Loss for encoding medical queries using BERT, (b) depicts the application of the resulting embeddings for multi-label intent classification, and (c) shows the enhancement of intent classification by integrating multi-source session context into the model.}
\label{fig:my_model_illustration}
\end{figure*}

\section{Methodology}
In this section, we first define the problem setting, and then  describe the fine-tuning and search intent prediction methodologies. 

\subsection{Problem Formulation}
In this study, we introduce an advanced approach for recognizing healthcare search query intent.  Our approach aims to fine-tune the search query embeddings using a novel multi-set loss function to align the query embeddings with the clicked document sets, cluster and classify the queries based on the learned embeddings, and further classify the queries in the context of search sessions. 
This section is methodically structured into three distinct parts, with two questions focusing on improving search query representation and one downstream task of search intent classification:  \\

\textbf{Query Representation Learning} Let $\mathcal{Q}$ be the set of all user queries and $\phi: \mathcal{Q} \rightarrow \mathbb{R}^d$ be the function mapping queries to their $d$-dimensional vector representations. Our goal is to learn the function $\phi$ such that for any pair of queries $(q, q') \in \mathcal{Q}$, the similarity in the embedding space reflects their semantic similarity.\\
This can be expressed as an optimization problem:
\begin{equation}
    \min_{\phi} \sum_{(q, q') \in \mathcal{Q}} \mathcal{L}\left(\phi(q), \phi(q')\right),
\end{equation}
where $\mathcal{L}$ is a loss function measuring the discrepancy between the embeddings of queries $q$ and $q'$, which should be small for semantically similar queries and large for dissimilar ones. The choice of $\mathcal{L}$ depends on the application's specific requirements. 

\textbf{Multi-label Search Intent Classification} Let $\mathcal{Q}$ be the set of all user queries and $\mathcal{Y}$ be the set of all potential intent labels, with $I$ being the total number of distinct intents, i.e., $I = |\mathcal{Y}|$. For each query $q \in \mathcal{Q}$, the task is to predict a subset of labels $\mathcal{Y}_q \subseteq \mathcal{Y}$ that accurately reflects the intents of $q$. This is formulated as a mapping function $f: \mathcal{Q} \rightarrow 2^{\mathcal{Y}}$, where $2^{\mathcal{Y}}$ is the power set of $\mathcal{Y}$. The query $q$ is represented by its embedding $\mathbf{E}(q)$, from which a probability distribution over the intent labels in $\mathcal{Y}$ is derived. Labels are then selected based on a thresholding mechanism to form the set $\mathcal{Y}_q$. The objective in training the model is to optimize a loss function that evaluates the accuracy of predicted labels for each query. This process typically involves techniques like Binary Cross-Entropy (BCE) applied across all $I$ labels. Labels are then selected based on a thresholding mechanism to form the set $\mathcal{Y}_q$. 

\subsection{Proposed Approach}
\textbf{RQ1: Enhancing Query Representation Learning}
As shown in Figure \ref{fig:my_model_illustration}a, our methodology utilizes the transformer-based query encoder (i.e. BERT \cite{devlin2018BERT}) as the initial query encoder. We enhance query representations through contrastive loss functions $\mathcal{L}_{contrast}$, tailoring the encoder to the intents of health search queries. The contrastive loss functions distinguish between pairs of queries clicked on the same document denoted as $\left(q, q^*\right)$ and those clicked on different documents denoted as $\left(q, q^{-}\right)$, ensuring the encoder captures this distinction. The representation of each query, ${E}_{i_j}$ for ${q}_{i_j}$ in the document set $\mathcal{C}_i$, is optimized to align closely with other queries in the document set $\mathcal{C}_i$ and diverge from dissimilar queries in the document set $\mathcal{C}_j$, with the total number of document sets denoted as $K$. This process refines our model to discern the semantic difference of search queries, informed by the user-clicked document annotations.

\textbf{RQ2: Enhancing Query Representation from Search Context} In RQ1, the query encoder is learned from the user behavior in aggregated search logs, and the search intents are inferred from the global statistics. However, the search intents could be different in different session contexts. To address this issue, we introduce the session context information to enhance the query representation for personalized search intent. Let $\mathcal{S} = {\mathbf{s}_1, \mathbf{s}_2, \ldots, \mathbf{s}_n}$ denote the set of session context vectors, each vector $\mathbf{s}_i$ capturing the temporal and relational dynamics of user interactions within a session. The session context includes the search queries $\mathbf{s}_n$, user clicked document annotation $\mathbf{a}_n$ and search page context $\mathbf{p}_n$ at search step $n$. Our process involves analyzing these vectors to predict the categorization of queries at step $n$. The representation of each query $q_i$ is optimized to align with the corresponding session context $\mathbf{s}_i$, thus enhancing the model's ability to tailor search results to individual user sessions. This methodology refines the query representation to capture user intent more accurately in session context.

\textbf{Downstream task: Multi-label Search Intent Classification.} In the task of multi-label search intent classification, we refine query embeddings and map them to the predefined categories using the classification loss funtion $\mathcal{L}_{class}$. Given the pre-trained query encoder from the previous step, our aim is to predict the probability distribution of each query $q$ across all predefined healthcare intent categories in $Y$. To achieve this, we transform the encoder's output into a probability vector, where each element corresponds to the likelihood of $q$ belonging to a specific category in $Y$. This probabilistic approach allows for a comprehensive understanding of $q$'s alignment with each potential intent category. The classification model, denoted as $\mathbf{M}$, takes the pre-trained query encoder and outputs a probability distribution over $\mathbf{Y}$, predicting the most likely category. This process is formalized as $\mathbf{y} = \mathbf{M}(\mathbf{E}(q))$, optimizing the mapping from query embeddings to category predictions.



\subsection{Pairwise Loss Function}
In this study, we focus on fine-tuning the embeddings of search queries to gain deeper insights into user intents. Following previous studies \cite{wang2022recognizing, zhang2019generic}, we first employ a pairwise loss function that optimizes the embeddings of the search queries based on their click patterns in a contrastive learning manner.

Document type refers to the category of a document within a search corpus. It is a metadata attribute that provides information about the origin or content of the document. Formally, let $\left(q, q^*\right)$ represent a pair of co-click queries on the same document type, and $\left(q, q^{-}\right)$ represent a pair of queries on different document types. The pairwise loss function, denoted as \(l_{\text{pairwise}}\), is defined as follows:
\vspace{-0.3cm}

\begin{equation}
\begin{split}
l_{\text{pairwise}} = \sum_q \bigg( & \frac{1}{1 + \exp\left(\cos\left(E(q), E(q^*)\right)\right)} \\
& - \frac{1}{1 + \exp\left(\cos\left(E(q), E(q^-)\right)\right)} \bigg)
\end{split}
\end{equation}


\subsection{Multiset Loss Function}
The multiset loss function is designed to mitigate noise in user co-click log data by utilizing a clustering-based approach. This method contrasts batches of search queries within the same cluster against those in different clusters, enhancing the accuracy of the representation learning process. The detailed derivation and mathematical formulation of the multiset loss function can be found in the supplemental material.

Central to the multiset loss function is the cosine similarity measure, which is computed between individual query embeddings and the centroid of embeddings corresponding to queries that share the same clicked document type. This approach combines intra-set and inter-set loss components to effectively capture the similarities and differences in query embeddings within and across different document sets. The detailed explanation for deriving the multiset loss function is provided in Section 4 of the Supplementary Material.

\textbf{Intra-Set Loss Function:} This component measures the average similarity of query embeddings within the same set. It is formulated as:
\begin{equation}
l_{\text{intra}} = \sum_{i=1}^{K}\frac{1}{N_i}\sum_{j=1}^{N_i} w_{i_j} \frac{1}{1 - \exp\left(\frac{\mathbf{E}_{i_j} \cdot \mathbf{C}_i}{\|\mathbf{E}_{i_j}\|\|\mathbf{C}_i\|}\right) / e + \epsilon}
\end{equation}
where $w_{i_j}$ weights the contribution of each query embedding $\mathbf{E}_{i_j}$ in cluster $i$, and $\epsilon$ is a small value added to prevent division by zero.

\textbf{Inter-Set Loss Function:} Conversely, the inter-set loss function is introduced to maximize the distance between embeddings of queries from different sets. This aspect is crucial for ensuring that queries from distinct sets are not mistakenly grouped together during representation learning. The inter-set loss is defined as:
\begin{equation}
l_{\text{inter}} = \sum_{i=1}^{K}\sum_{\substack{j=1\\j \neq i}}^{K}\frac{1}{N_i}\sum_{k=1}^{N_i} {w_{i_k}}  \frac{1}{1 - \exp\left(\frac{\mathbf{E}_{i_k} \cdot \mathbf{C}_j}{\|\mathbf{E}_{i_k}\|\|\mathbf{C}_j\|}\right) / e + \epsilon}
\end{equation}
where $w_{i_k}$ weights the contribution of each query embedding $\mathbf{E}_{i_k}$ in cluster $i$, and $\epsilon$ is a small value added to prevent division by zero.

\textbf{Multiset Loss Function:} The multiset loss function $l_{\text{multiset}}$ elegantly combines these two components, encapsulating our dual objectives of enhancing intra-cluster cohesion and inter-cluster separation:
\begin{equation}
\label{eq:weighted multiset loss}
l_{\text{multiset}} = -\log\left(\frac{l_{\text{intra}}}{l_{\text{inter}}}\right)
\end{equation}

For more details, refer to the supplemental material.

\subsection{Multi-Label Search Intent Classification}
\label{sec:multi-label-search-classification}
In this study, we treat multi-label search intent classification as our major downstream task after query representation learning. The objective of multi-label intent classification is to assign a set of intent labels to a given input text. 


\textbf{Model Architecture} Our model architecture is based on the Transformer-based BERT model. Given the pre-trained BERT model, we further fine-tune the model on the downstream intent classification dataset using a custom loss function designed for multi-label classification. We employ our pre-trained MSet-BERT to obtain dense vector representations for each search query. For the sequence of search query terms, we get the final vector corresponding to the [CLS] token, which is designed to hold the aggregate sequence representation, and is then passed through a series of fully connected layers for each search intent.

\textbf{Multi-Label Loss Function}
For training our model, we use the Binary Cross-Entropy (BCE) loss, treating each label as an independent binary classification. The BCE loss for a single instance $(x, Y)$ with the predicted output $\hat{Y}$ is defined as:

\begin{equation}
\text{BCE}(Y, \hat{Y}) = -\frac{1}{|\mathcal{C}|} \sum_{i \in \mathcal{C}} [y_i \log(\hat{y}_i) + (1 - y_i) \log(1 - \hat{y}_i)]
\end{equation}

where $y_i$ is the binary indicator (0 or 1) if label $i$ is the correct classification for $x$, and $\hat{y}_i$ is the predicted probability of $x$ having label $i$.


\textbf{Training and Evaluation}
As shown in Table \ref{tab:train_test_atasets}, we split our datasets into training, validation, and test sets and evaluated our model. The ambiguity of search query is measured using Perplexity. The performance is measured using appropriate multi-label metrics, including Precision, and F1 score. In addition, we also order the probability of the intents and evaluate the rank using metrics including Hit Rate@3 and NDCG@3 for multiple intents retrieval. The detailed formula for the training and evaluation metrics is provided in Section 5 of the Supplementary Material.

\subsection{Contextual Search Intent Classification}
To enhance the search representation in sessions with different concordance rates, we aim to predict the intent of a search query within the context of previous user interactions in the same session. This approach considers a series of queries and actions to understand their unique information goals. 

\textbf{Concordance Rate} The concordance rate is a metric designed to quantify the alignment between the intents of queries within search sessions and the intents inferred from unique search queries based on search logs. It serves as an indicator of the difficulty level in predicting search session intent, as it reflects the extent to which intents within a session deviate from the standard patterns recognized by pre-trained models.\\

\begin{equation}
\label{eq:cr_rate}
    CR(S_i) = 
    \frac{1}{|S_i|} \sum_{q \in S_i} \mathbb{I}(\text{Intent}(q) = 
    \text{Session-Inferred Intent}(q))
\end{equation}

Here, $\mathbb{I}$ is the indicator function that returns 1 if the intents match and 0 otherwise, $|S_i|$ is the number of queries in session $S_i$, $\text{Intent}(q)$ is the global intent of query $q$ from search logs, and $\text{Session-Inferred Intent}(q)$ is the intent based on the session clicks. A lower concordance rate indicates a session with high difficulty, suggesting that the user's intent is not well-aligned with the patterns learned by the search logs. \\
\textbf{Model Achitecture} Our model extends the BERT architecture by incorporating a multi-source cross attention mechanism that integrates various sources of session context as described in \cite{hashemi2020guided}. The architecture is designed to capture the nuances of each user interaction until the current search query. The input of the multi-source cross attention model is the concatenation of the current search query, previous search queries, previous clicked document annotations, and page context into a single sequence. Special tokens [CLS] and [SEP] are used to separate different elements:
\begin{align}
    &\text{Input} = [\text{CLS}] \; q_n \; [\text{SEP}] \; q_{n-1} \; [\text{SEP}] \; ... \; [\text{SEP}] \; a_{n-1} \\
    &\; [\text{SEP}] \; ... \; [\text{SEP}] \; p_{n-1}
\end{align}
where $p_{n-1}$ denotes the page context associated with the $(n-1)^{th}$ query.


\begin{table}[t!]
\vspace{-0.3cm}
\centering
\caption{Distribution of the train, validation, and test datasets for the HS and TripClick datasets.}
\vspace{-0.3cm}
\begin{tabular}{ccc}
\toprule
                     & \textbf{HS} & \textbf{TripClick} \\
\midrule
\textbf{Train dataset} & 27,170             & 81,222                \\
\textbf{Validation dataset}   & 9,056            & 27,078               \\
\textbf{Test dataset}  & 9,060            & 32,044                  \\
\bottomrule
\end{tabular}

\label{tab:train_test_atasets}
\end{table}

\section{Experiments}
In this section, we will first describe the datasets used in this study, then introduce the experimental setup for the query intent classification task and the session-based search intent classification task.
\subsection{Datasets}

\paragraph{Health Search (HS) Dataset}
The HS dataset, collected from a health website internal search engine, spans Jan 2022 to Sept 2023. It includes queries, clicked documents, query-document pairs, and document attributes like titles, document URLs, and document types. The detailed dataset descriptions are provided in Section 1.1 of the Supplementary Material.

\paragraph{TripClick Dataset}
TripClick is a large-scale dataset of click logs in the health domain from the Trip Database health web search engine. The dataset contains about 5.2 million user interactions collected between 2013 and 2020. The detailed dataset descriptions are provided in Section 1.2 of the Supplementary Material.

 \subsection{Experimental Setup for Intent Classification}
 Understanding the user intent is key to personalized healthcare experiences. This requires a meticulous approach, first defining a comprehensive query intent taxonomy and then accurately annotating the queries into query intents, as shown in Figure \ref{fig:label_distributions}. We describe the method for extracting the queries using co-query set extractor in Section 2 of the Supplementary Material. The detailed methodology for intent annotation is provided in Section 3 of the Supplementary Material.
 
 We create the train, validation, and test sets for intent classification using a variation of stratified sampling in approximately 60:20:20 ratio. Since we perform multi-label classification, we first group the queries by label combinations. We then divide the samples from each group among the three sets.
 
 For the HS dataset, all sets have the same proportion of class labels/class label combinations. Their sizes are in the ratio 60:20:20. All three sets have similar distributions over class labels. 

 The TripClick dataset has a larger number of classes (document types) and some combinations of labels are extremely infrequent. For simplicity, we first create the train-validation-test split with labels that have at least 3 samples. We then add all the queries with unique label combinations to the test set. 

 \subsection{Experimental Setup for Session-based Intent Classification}
For session-based intent classification, we carefully curate search sessions based on their sequence lengths, specifically ranging from 2 to 6 and session length of 4, as per \cite{ortiz2022exploring} suggested, was used for representing contextual search. This range is chosen to effectively separate the datasets for training, validation, and testing, ensuring each set represents varying complexities of user interactions. For the training set, we utilize sessions from search step 1 to search step n-1, focusing on building the training model. The validation and test sets comprise sessions from search step 2 to search step n, aiming to determine the classification threshold and evaluate the model’s performance. The specific criteria for session selection, rooted in sequence length, are designed to represent real-world user search behavior, thus enhancing the relevance and implication of the classification task.

\subsection{Baseline Models}
\paragraph{BERT} BERT model \cite{devlin2018BERT} is a pre-trained language model trained on the large-scale corpus. It can be fine-tuned on the downstream tasks via either the contrastive loss functions or classification loss functions. The BERT base model (uncased) is used as one of the baseline models and tailored to our study's objectives. 

\paragraph{PairWise-BERT} The PairLoss-BERT model represents an advanced baseline \cite{zhang2019generic, wang2022recognizing}. It undergoes pre-training with a contrastive learning approach using the pairwise loss function. This model sets a state-of-the-art baseline for our approach, demonstrating the effectiveness of contrastive learning in query representation.

\paragraph{MSet-BERT (our method) with variants for ablation studies}
We report performance of several variants of the MSet-BERT model to assess the impact of different session contexts in session-based search intent classification:
\begin{itemize}
    \item \textbf{BERT w all context (Ablation):} Enhances the baseline BERT model by incorporating complete session context.
    \item \textbf{PairWise-BERT w all context (Ablation):} Builds upon the PairWise-BERT model by integrating full session context.
    \item \textbf{MSet-BERT w/o context:} Uses only the search queries as input, serving as a baseline to understand the model's performance without any session context.
    \item \textbf{MSet-BERT w prev-query:} Includes the previous query in the session as part of the context.
    \item \textbf{MSet-BERT w page-context:} Incorporates the context of the web page the user search happens.
\end{itemize}

\begin{table}[ht!]
\centering
\caption{Comparative clustering performance analysis of query representations from BERT, PairWise-BERT, and MSet-BERT models on HS and TripClick test datasets using adjusted rand index (ARI) and normalized mutual information (NMI). The best performing values are highlighted in bold.}
\small 
\setlength{\tabcolsep}{3pt} 
\begin{tabular}{lcccc}
\toprule
\small
\textbf{Dataset} & \textbf{Models} & \textbf{ARI } & \textbf{NMI} \\

\midrule
\multirow{3}{*}{HS} & BERT & 0.0540 & 0.0859 \\
& PairWise-BERT & 0.0690 & 0.1161 \\
& {\bf MSet-BERT} & \textbf{0.0723} (+4.78\%) & \textbf{0.1181} (+1.72\%) \\
\hline
\multirow{3}{*}{TripClick} & BERT & 0.00189 & 0.0120 \\
& PairWise-BERT & 0.00213 & 0.0113 \\
& {\bf MSet-BERT} & \textbf{0.00235}  (+10.33\%)& \textbf{0.0122} (+1.67\%) \\
\bottomrule

\end{tabular}
\label{tab:clustering_comparison}
\end{table}

\begin{table*}[ht!]
\centering
\small
\caption{Evaluation of model performance on the HS and TripClick datasets across multiple metrics: Precision, F1, Hit Rate@3, and NDCG@3. The highest performing values are marked in bold, with asterisks denoting significant improvements over the best baseline model. Statistical significance is determined by a t-test for \(N\) queries in the test dataset, with \(p < 0.05\).}
\begin{tabular}{lccccc}
\toprule
\textbf{Dataset} & \textbf{Model} & \textbf{Precision} & \textbf{F1} & \textbf{Hit Rate@3} & \textbf{NDCG@3} \\

\midrule
\multirow{3}{*}{HS Dataset} 
& BERT & 0.937 & 0.932 & 0.976 & 0.937 \\
& PairWise-BERT & 0.943 & 0.936 & 0.988 & 0.940 \\
& {\bf MSet-BERT} & \textbf{0.970*} (+2.86\%) & \textbf{0.969*} (+3.53\%) & \textbf{0.992*} (+0.40\%) & \textbf{0.975*} (+3.72\%) \\
\hline
\multirow{3}{*}{TripClick Dataset} 
& BERT & 0.870 & 0.813 & 0.947 & 0.853 \\
& PairWise-BERT & 0.881 & 0.840 & 0.956 & 0.875 \\
& {\bf MSet-BERT} & \textbf{0.895*} (+1.59\%) & \textbf{0.854*} (+1.67\%) & \textbf{0.965*} (+0.94\%) & \textbf{0.886*} (+1.26\%) \\
\bottomrule

\end{tabular}
\label{tab:model_classification_comparison}
\end{table*}

\begin{table*}[ht!]
\centering
\small
\caption{Comparative performance of MSet-BERT models in session-based intent classification on the HS and TripClick datasets, highlighting the impact of different context integration strategies (no context, previous query, page context, and all contexts). Performance metrics include Precision, F1, Hit Rate@3, and NDCG@3, with the best scores highlighted in bold and marked with an asterisk (*) to indicate significant improvement. Statistical significance is determined by a t-test for \(N\) queries in the test dataset, with \(p < 0.05\). Session length of 4, as per \cite{ortiz2022exploring}, was used for this analysis.}
\begin{tabular}{lccccc}
\toprule

\textbf{Dataset (Concordance)} & \textbf{Model (Ablation)}  & \textbf{Precision} & \textbf{F1} & \textbf{Hit Rate@3} & \textbf{NDCG@3} \\
\midrule
\multirow{6}{*}{HS (33\%)} 
& BERT w all context (Abl) & 0.715 & 0.722  & 0.856 & 0.658  \\
& PairWise-BERT w all context (Abl) & 0.736 & 0.749  & 0.866 & 0.678  \\
& MSet-BERT w/o context & 0.707 & 0.708  & 0.864 & 0.653 \\
& MSet-BERT w prev-query & 0.721 & 0.753  & 0.871 & 0.669 \\
& MSet-BERT w page-context & 0.759 & 0.751  & 0.869 & 0.667 \\
    & {\bf MSet-BERT w all context} & \textbf{0.776*} (+5.44\%) & \textbf{0.784*} (+4.67\%)  & \textbf{0.880*} (+1.61\%)  & \textbf{0.769*} (+13.42\%) \\
\hline
\multirow{5}{*}{TripClick (88\%)}
& BERT w all context (Abl) & 0.829 & 0.814  & 0.931 & 0.765  \\
& PairWise-BERT w all context (Abl) & 0.846 & 0.820  & 0.936 & 0.772  \\
& MSet-BERT w/o context & 0.831 & 0.813  & 0.929 & 0.762 \\
& MSet-BERT w prev-query & 0.844 & 0.824  & 0.941 & 0.778 \\
    & {\bf MSet-BERT w all context} & \textbf{0.868*} (+2.60\%) & \textbf{0.840*} (+2.44\%)  & \textbf{0.953*} (+1.81\%) & \textbf{0.826*} (+7.00\%)  \\
\bottomrule
\end{tabular}
\label{tab:session_multiset_BERT_performance}
\end{table*}

\begin{figure}[htbp]
  \centering
    \centering
    \includegraphics[width=0.85\linewidth]{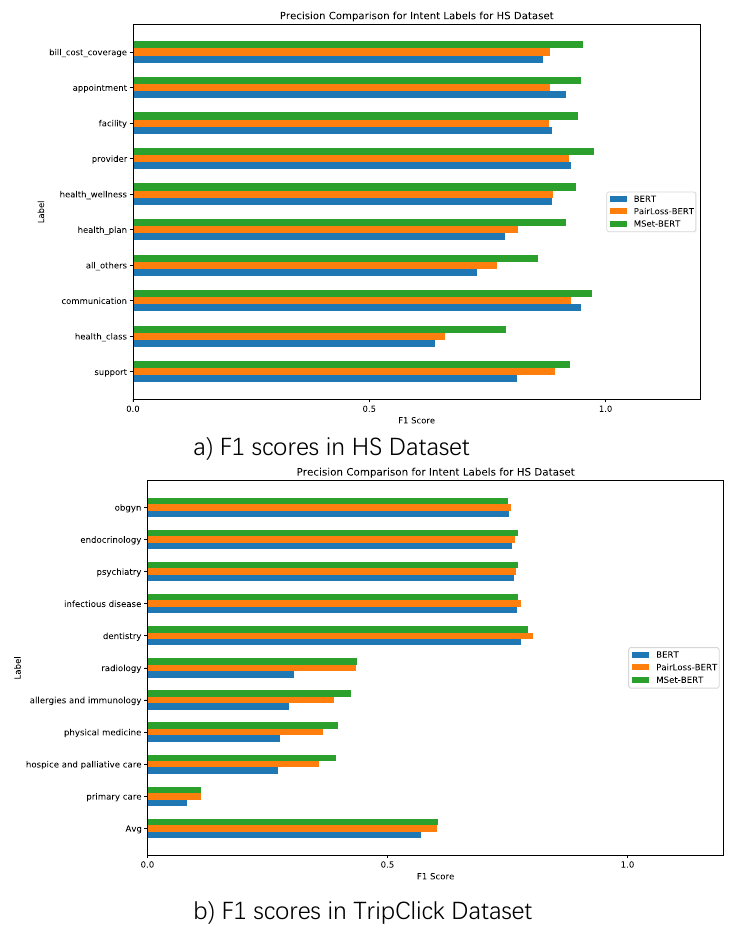}
    \vspace{-0.1cm}
  
  \caption{Comparative analysis of F1 scores for different intent types within the HS and TripClick dataset, providing insights into the model's performance in accurately classifying and retrieving relevant search intents.}
  \label{fig:intent_distributions}
\end{figure}



\begin{figure}[ht]
    \centering
    \includegraphics[width=\linewidth]{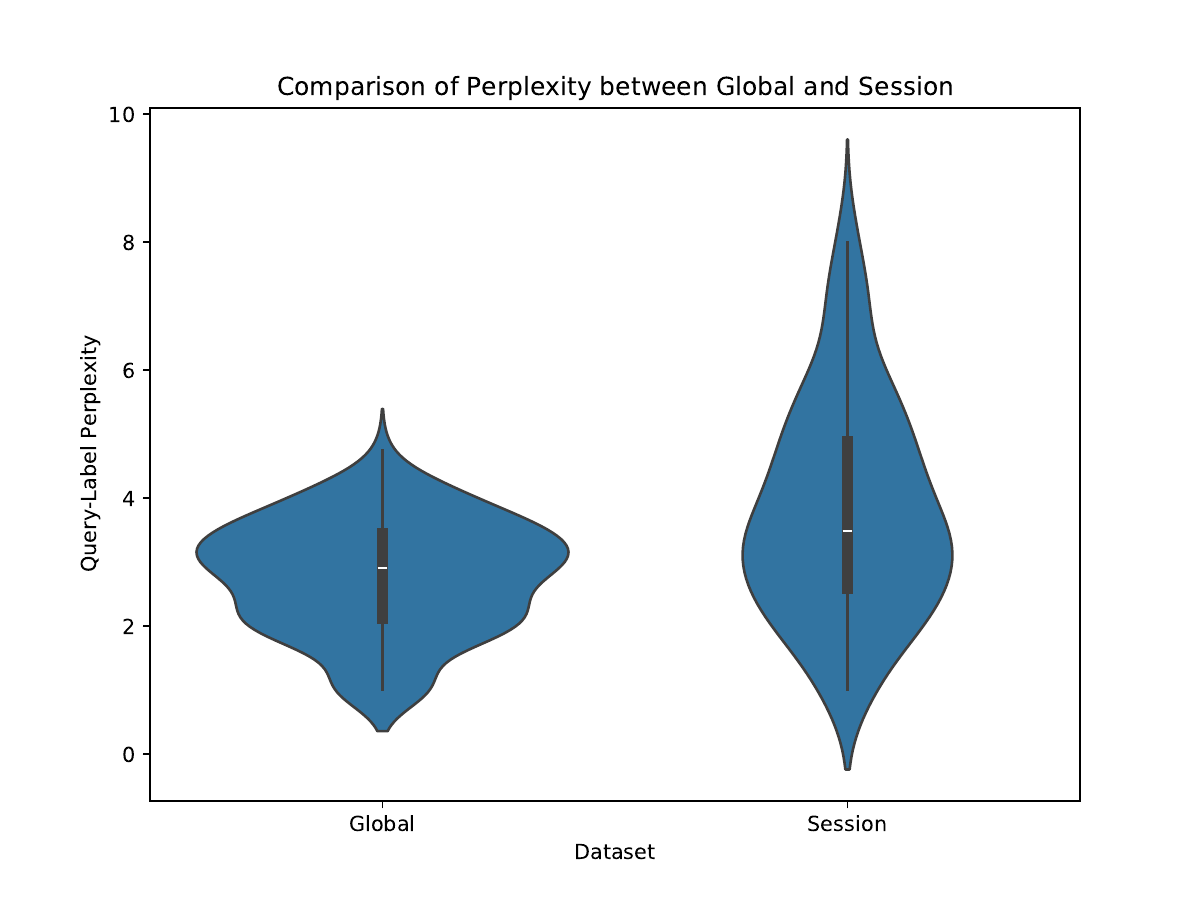}
    \vspace{-0.5cm}
    \caption{Comparison of query perplexity for the global and session-specific intent classification of 140 common queries in the HS dataset.}
    \label{fig:comp_perplexity}
\end{figure}

\begin{figure}[ht]
    \centering
    \includegraphics[width=\linewidth]{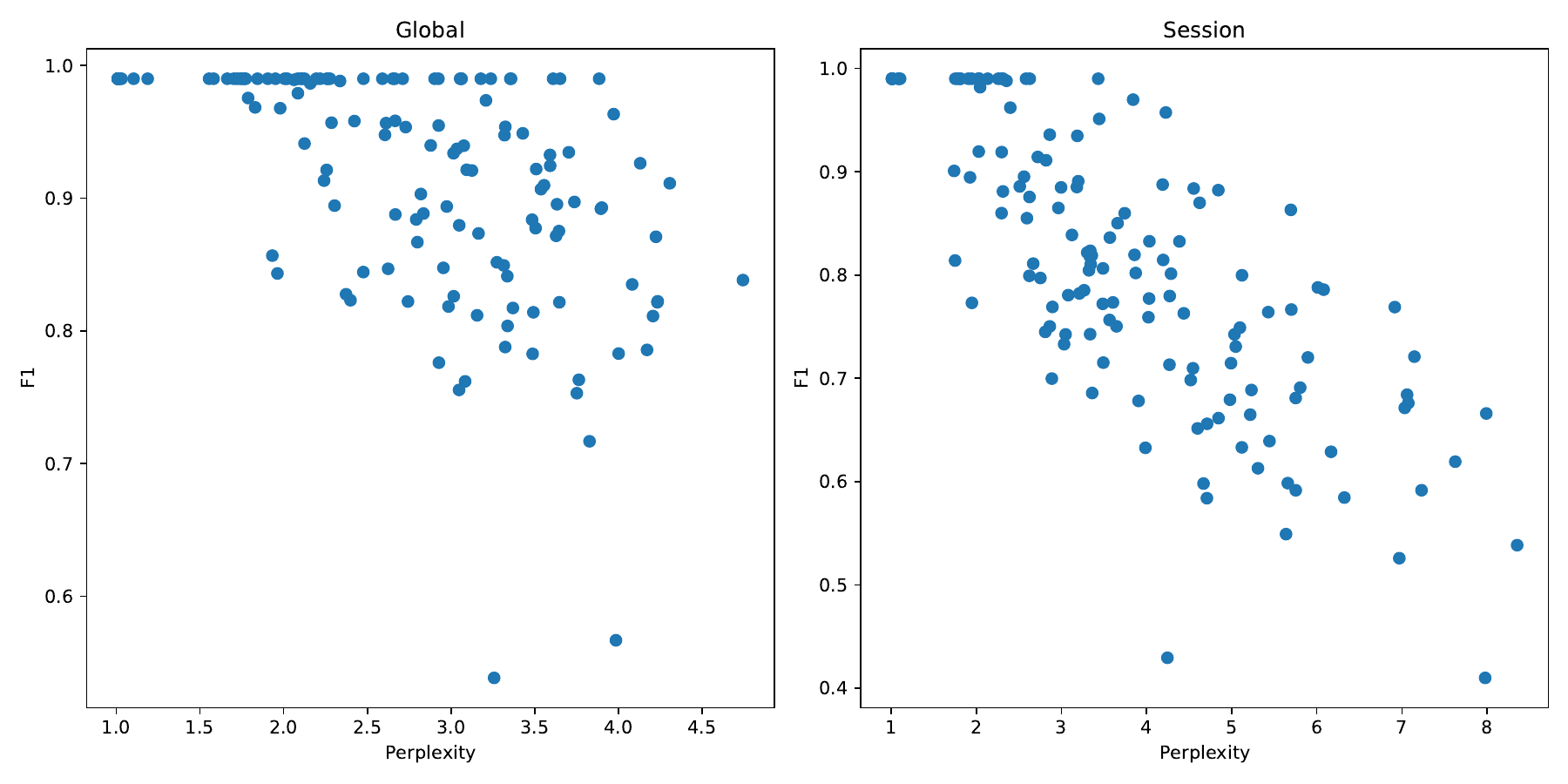}
        \vspace{-0.5cm}
    \caption{Comparison of query perplexity and  F1 scores for the global and session-specific intent classification of 140 common queries in the HS dataset.}
\label{fig:comp_f1_perplexity}
\end{figure}

\section{Result and Discussion}
\label{sec:results_discussion}
In this section, we describe the performance analysis of the MSet-BERT model, underscoring its advancements in multi-label search intent classification and session-based search intent classification over baseline models.

\subsection{Performance Comparison}
\label{subsec:performance_comparison}
In our study of the MSet-BERT model for multi-label search intent classification and session-based search intent classification, we observed significant improvements over baseline models. In the evaluation of query representation clustering, as detailed in Table \ref{tab:clustering_comparison}, the MSet-BERT model consistently outperforms the BERT and PairWise-BERT model on the ARI and NMI score for query representation clustering. Specifically, for the HS dataset, MSet-BERT achieves a 4.78\% higher ARI and a 1.72\% improvement in NMI compared to the best baseline. This trend is similarly observed in the TripClick dataset, where MSet-BERT shows a substantial improvement of 10.33\% in ARI and 1.67\% in NMI scores over the other models. These results underscore the efficacy of MSet-BERT in clustering query representations from search logs to match users' search intent.
In the evaluation of the downstream intent classification task, the MSet-BERT model also demonstrate significant improvement over the baseline models. Specifically, as in detailed Table \ref{tab:model_classification_comparison} and Figure \ref{fig:intent_distributions}, MSet-BERT demonstrates an improvement of 2.86\% in precision, 3.53\% in F1 score, 0.40\% in Hit Rate@3, and a notable 3.72\% in NDCG@3 for HS dataset. Simlarly, for the TripClick dataset, the model exhibits improvements of 1.59\% in precision, 1.67\% in F1 score, 0.94\% in Hit Rate@3, and 1.26\% in NDCG@3 compared to the best performing baseline model. These improvements underscore the effectiveness of MSet-BERT in enhancing the robustness and accuracy of search intent classification for health-related queries.
\subsection{Session-based Intent Classification Performance}
In session-based intent classification, predicting user search intents is a more difficult task because it requires the model to handle the situations where the user search behavior deviates or disagrees with the search-query-based intent. In the evaluation of MSet-BERT in the session-based search intent classification task, as described in Table \ref{tab:session_multiset_BERT_performance}, the MSet-BERT model demonstrates notable effectiveness compared with all baseline models. Specifically, the MSet-BERT model exhibits a marked improvement in precision (up to 5.44\% for HS and 2.60\% for TripClick), F1 score (increased by 4.67\% for HS and 2.44\% for TripClick), Hit Rate@3 (improved by 1.61\% for HS and 1.81\% for TripClick), and NDCG@3 (a substantial gain of 13.42\% for HS and 7.00\% for TripClick). The ablation studies without context or with only a part of session contexts also underscore the effectiveness of the MSet-BERT in harnessing context integration strategies for session-based intent 
classification.
\subsection{Ambiguity and Difficulty in Session-based Intent Classification}
A core challenge in session-based intent classification is recognizing and addressing the ambiguity of intent for the same queries across different sessions. As described in Figure \ref{fig:comp_perplexity}, the same queries when collected for global query intent (derived from the number of global co-occurrence in multiple sessions) and for session-based search intent (defined as the subsequent action in a single session) can exhibit different query intent perplexity, where higher perplexity means higher ambiguity. Further analysis, as shown in Figure \ref{fig:comp_f1_perplexity}, indicates a notable inverse correlation between the query perplexity and F1 scores, suggesting greater difficulty in classification. This implies that accurately identifying intents in session-based queries involves not only considering global intents but also intricately analyzing the context of each session.


\section{Conclusions}
In this paper, we proposed a novel method for learning query representation, focusing on the Health domain, resulting in an effective query embedding model MSet-BERT. The method proposed is general and uses a novel multiset loss function designed to capture the inherent ambiguity of health search queries, which results in an enhanced search query representation. Our MSet-BERT model, introduced, demonstrated its advantages over prior open SOTA models of BERT and PairWise-BERT models trained on search logs, for both intrinsic query clustering tasks, and for multi-label intent classification tasks. 

Furthermore, this study investigated the effectiveness of different context representations to improve session-based intent prediction, demonstrating that the MSet-BERT model generalizes to the more challenging individual session-based intent prediction task, even when the individual search intent disagrees with the most popular intent for the query based on global statistics. Our experiments show that adding context information from previous queries and clicked documents can improve the performance of the MSet-BERT model for the search session query intent recognition.

\bibliographystyle{ACM-Reference-Format}
\bibliography{main}

@inproceedings{zhang2019generic,
  title={Generic intent representation in web search},
  author={Zhang, Hongfei and Song, Xia and Xiong, Chenyan and Rosset, Corby and Bennett, Paul N and Craswell, Nick and Tiwary, Saurabh},
  booktitle={Proceedings of the 42nd International ACM SIGIR Conference on Research and Development in Information Retrieval},
  pages={65--74},
  year={2019}
}

@inproceedings{wang2022recognizing,
  title={Recognizing medical search query intent by few-shot learning},
  author={Wang, Yaqing and Wang, Song and Li, Yanyan and Dou, Dejing},
  booktitle={Proceedings of the 45th International ACM SIGIR Conference on Research and Development in Information Retrieval},
  pages={502--512},
  year={2022}
}

@inproceedings{agichtein2006improving,
  title={Improving web search ranking by incorporating user behavior information},
  author={Agichtein, Eugene and Brill, Eric and Dumais, Susan},
  booktitle={Proceedings of the 29th annual international ACM SIGIR conference on Research and development in information retrieval},
  pages={19--26},
  year={2006}
}

@inproceedings{rekabsaz2021tripclick,
  title={Tripclick: the log files of a large health web search engine},
  author={Rekabsaz, Navid and Lesota, Oleg and Schedl, Markus and Brassey, Jon and Eickhoff, Carsten},
  booktitle={Proceedings of the 44th International ACM SIGIR Conference on Research and Development in Information Retrieval},
  pages={2507--2513},
  year={2021}
}

@article{devlin2018bert,
  title={Bert: Pre-training of deep bidirectional transformers for language understanding},
  author={Devlin, Jacob and Chang, Ming-Wei and Lee, Kenton and Toutanova, Kristina},
  journal={arXiv preprint arXiv:1810.04805},
  year={2018}
}

@article{lee2020biobert,
  title={BioBERT: a pre-trained biomedical language representation model for biomedical text mining},
  author={Lee, Jinhyuk and Yoon, Wonjin and Kim, Sungdong and Kim, Donghyeon and Kim, Sunkyu and So, Chan Ho and Kang, Jaewoo},
  journal={Bioinformatics},
  volume={36},
  number={4},
  pages={1234--1240},
  year={2020},
  publisher={Oxford University Press}
}

@article{rahman2013search,
  title={Search engines going beyond keyword search: a survey},
  author={Rahman, Mahmudur},
  journal={Int. J. Comput. Appl},
  volume={75},
  number={17},
  pages={1--8},
  year={2013}
}

@inproceedings{broder2002taxonomy,
  title={A taxonomy of web search},
  author={Broder, Andrei},
  booktitle={ACM Sigir forum},
  volume={36},
  number={2},
  pages={3--10},
  year={2002},
  organization={ACM New York, NY, USA}
}

@inproceedings{broder2007robust,
  title={Robust classification of rare queries using web knowledge},
  author={Broder, Andrei Z and Fontoura, Marcus and Gabrilovich, Evgeniy and Joshi, Amruta and Josifovski, Vanja and Zhang, Tong},
  booktitle={Proceedings of the 30th annual international ACM SIGIR conference on Research and development in information retrieval},
  pages={231--238},
  year={2007}
}

@inproceedings{yin2010building,
  title={Building taxonomy of web search intents for name entity queries},
  author={Yin, Xiaoxin and Shah, Sarthak},
  booktitle={Proceedings of the 19th international conference on World wide web},
  pages={1001--1010},
  year={2010}
}

@inproceedings{xiong2017end,
  title={End-to-end neural ad-hoc ranking with kernel pooling},
  author={Xiong, Chenyan and Dai, Zhuyun and Callan, Jamie and Liu, Zhiyuan and Power, Russell},
  booktitle={Proceedings of the 40th International ACM SIGIR conference on research and development in information retrieval},
  pages={55--64},
  year={2017}
}

@inproceedings{hu2009understanding,
  title={Understanding user's query intent with wikipedia},
  author={Hu, Jian and Wang, Gang and Lochovsky, Fred and Sun, Jian-tao and Chen, Zheng},
  booktitle={Proceedings of the 18th international conference on World wide web},
  pages={471--480},
  year={2009}
}

@inproceedings{nalisnick2016improving,
  title={Improving document ranking with dual word embeddings},
  author={Nalisnick, Eric and Mitra, Bhaskar and Craswell, Nick and Caruana, Rich},
  booktitle={Proceedings of the 25th International Conference Companion on World Wide Web},
  pages={83--84},
  year={2016}
}

@inproceedings{zamani2017relevance,
  title={Relevance-based word embedding},
  author={Zamani, Hamed and Croft, W Bruce},
  booktitle={Proceedings of the 40th International ACM SIGIR Conference on Research and Development in Information Retrieval},
  pages={505--514},
  year={2017}
}

@article{cer2018universal,
  title={Universal sentence encoder},
  author={Cer, Daniel and Yang, Yinfei and Kong, Sheng-yi and Hua, Nan and Limtiaco, Nicole and John, Rhomni St and Constant, Noah and Guajardo-Cespedes, Mario and Yuan, Steve and Tar, Chris and others},
  journal={arXiv preprint arXiv:1803.11175},
  year={2018}
}

@inproceedings{hashemi2020guided,
  title={Guided transformer: Leveraging multiple external sources for representation learning in conversational search},
  author={Hashemi, Helia and Zamani, Hamed and Croft, W Bruce},
  booktitle={Proceedings of the 43rd international acm sigir conference on research and development in information retrieval},
  pages={1131--1140},
  year={2020}
}

@inproceedings{ortiz2022exploring,
  title={Exploring the Value of Multi-View Learning for Session-Aware Query Representation},
  author={Ortiz, Diego and Moreno, Jos{\'e} G and Hubert, Gilles and Pinel-Sauvagnat, Karen and Tamine, Lynda},
  booktitle={Annual Conference of the North American Chapter of the Association for Computational Linguistics (NAACL 2022)},
  pages={304--315},
  year={2022},
  organization={ACL: Association for Computational Linguistics}
}

@inproceedings{yuan2023multi,
  title={A Multi-Granularity Matching Attention Network for Query Intent Classification in E-commerce Retrieval},
  author={Yuan, Chunyuan and Qiu, Yiming and Li, Mingming and Hu, Haiqing and Wang, Songlin and Xu, Sulong},
  booktitle={Companion Proceedings of the ACM Web Conference 2023},
  pages={416--420},
  year={2023}
}

@String{Computing = "Computing" }

@String{Computer = "{IEEE} Computer" }

@ArtifactSoftware{R,
    title = {R: A Language and Environment for Statistical Computing},
    author = {{R Core Team}},
    organization = {R Foundation for Statistical Computing},
    address = {Vienna, Austria},
    year = {2019},
    url = {https://www.R-project.org/},
}

@inproceedings{schroff2015facenet,
  title={Facenet: A unified embedding for face recognition and clustering},
  author={Schroff, Florian and Kalenichenko, Dmitry and Philbin, James},
  booktitle={Proceedings of the IEEE conference on computer vision and pattern recognition},
  pages={815--823},
  year={2015}
}

@inproceedings{shen2014learning,
  title={Learning semantic representations using convolutional neural networks for web search},
  author={Shen, Yelong and He, Xiaodong and Gao, Jianfeng and Deng, Li and Mesnil, Gr{\'e}goire},
  booktitle={Proceedings of the 23rd international conference on world wide web},
  pages={373--374},
  year={2014}
}

@inproceedings{jansen2007determining,
  title={Determining the user intent of web search engine queries},
  author={Jansen, Bernard J and Booth, Danielle L and Spink, Amanda},
  booktitle={Proceedings of the 16th international conference on World Wide Web},
  pages={1149--1150},
  year={2007}
}

@inproceedings{Matt:2013,
  title        = {Explicit and implicit syntactic features for text classification},
  author       = {Post, Matt and Shane Bergsma.},
  booktitle    = {proceedings of the 51st Annual Meeting of the Association for Computational Linguistics},
  pages        = {866-872},
  year         = {2013},
  organization = {}
}

@inproceedings{wang2017combining,
  title={Combining Knowledge with Deep Convolutional Neural Networks for Short Text Classification.},
  author={Wang, Jin and Wang, Zhongyuan and Zhang, Dawei and Yan, Jun},
  booktitle={IJCAI},
  volume={350},
  pages={3172077--3172295},
  year={2017}
}

@inproceedings{wang2015query,
  title={Query understanding through knowledge-based conceptualization},
  author={Wang, Zhongyuan and Zhao, Kejun and Wang, Haixun and Meng, Xiaofeng and Wen, Ji-Rong},
  booktitle={IJCAI},
  year={2015}
}

@inproceedings{white2010predicting,
  title={Predicting short-term interests using activity-based search context},
  author={White, Ryen W and Bennett, Paul N and Dumais, Susan T},
  booktitle={Proceedings of the 19th ACM international conference on Information and knowledge management},
  pages={1009--1018},
  year={2010}
}

@inproceedings{bennett2012modeling,
  title={Modeling the impact of short-and long-term behavior on search personalization},
  author={Bennett, Paul N and White, Ryen W and Chu, Wei and Dumais, Susan T and Bailey, Peter and Borisyuk, Fedor and Cui, Xiaoyuan},
  booktitle={Proceedings of the 35th international ACM SIGIR conference on Research and development in information retrieval},
  pages={185--194},
  year={2012}
}

@inproceedings{hashemi2021learning,
  title={Learning multiple intent representations for search queries},
  author={Hashemi, Helia and Zamani, Hamed and Croft, W Bruce},
  booktitle={Proceedings of the 30th ACM International Conference on Information \& Knowledge Management},
  pages={669--679},
  year={2021}
}

@article{srinivasan2022quill,
  title={QUILL: Query intent with large language models using retrieval augmentation and multi-stage distillation},
  author={Srinivasan, Krishna and Raman, Karthik and Samanta, Anupam and Liao, Lingrui and Bertelli, Luca and Bendersky, Mike},
  journal={arXiv preprint arXiv:2210.15718},
  year={2022}
}

@inproceedings{shen2006building,
  title={Building bridges for web query classification},
  author={Shen, Dou and Sun, Jian-Tao and Yang, Qiang and Chen, Zheng},
  booktitle={Proceedings of the 29th annual international ACM SIGIR conference on Research and development in information retrieval},
  pages={131--138},
  year={2006}
}

@inproceedings{kong2015predicting,
  title={Predicting search intent based on pre-search context},
  author={Kong, Weize and Li, Rui and Luo, Jie and Zhang, Aston and Chang, Yi and Allan, James},
  booktitle={Proceedings of the 38th International ACM SIGIR Conference on Research and Development in Information Retrieval},
  pages={503--512},
  year={2015}
}

@inproceedings{zamani2017situational,
  title={Situational context for ranking in personal search},
  author={Zamani, Hamed and Bendersky, Michael and Wang, Xuanhui and Zhang, Mingyang},
  booktitle={Proceedings of the 26th International Conference on World Wide Web},
  pages={1531--1540},
  year={2017}
}

@article{sen2021know,
  title={I know what you need: Investigating document retrieval effectiveness with partial session contexts},
  author={Sen, Procheta and Ganguly, Debasis and Jones, Gareth JF},
  journal={ACM Transactions on Information Systems (TOIS)},
  volume={40},
  number={3},
  pages={1--30},
  year={2021},
  publisher={ACM New York, NY}
}

@article{vuong2024predicting,
  title={Predicting Representations of Information Needs from Digital Activity Context},
  author={Vuong, Tung and Ruotsalo, Tuukka},
  journal={ACM Transactions on Information Systems},
  year={2024},
  publisher={ACM New York, NY}
}





\setcounter{section}{0}
\clearpage
\onecolumn
\pagenumbering{gobble}

\begin{center}
    {\LARGE \textbf{Supplementary Material}}
\end{center}

\pagenumbering{arabic}

\section{Detailed Dataset Descriptions}

\subsection{Health Search (HS) Dataset}
The HS dataset is the internal dataset from the search engine of Company A. The HS dataset contains about 225 thousand search queries, and 18 thousand clicked URLs. The dataset comprises search logs from January 2022 to November 2022. The HS dataset contains the following information: the queries, the clicked documents, the query-document pairs, and document attributes. The document attributes include the document title, the document URL, and the document type (i.e. label). 

When we generate user-clicked data, we filter the click logs by the click count greater than 2. For session-based search intent classification, we specifically set aside the data from January 2022, which is not included in the representation learning dataset. This exclusion is methodologically significant as it allows for a focused comparison between global and session-level search intents using a distinct and controlled dataset. January 2022 was chosen for this analysis to provide a clear baseline for assessing how individual search intents align or diverge over global search intent, thus offering insights into the user search behavior in healthcare contexts.

For evaluation purposes, with the help of HS experts, we independently built the query intent clustering and classification dataset as shown in Table 1. The dataset is generated by expert knowledge, the queries are manually labeled into one of the intent queries from ["access\_records", "account\_mgmt", "appointment", "bill\_cost\_coverage", "communication", "drug\_info", "health\_wellness", "provider", "facility", "health\_class", "health\_plan", "job\_search", "support", "all\_others"].

\subsection{TripClick Dataset}
TripClick is a large-scale dataset of click logs in the health domain from the Trip Database health web search engine \cite{rekabsaz2021tripclick}. The TripClick dataset contains about 5.2 million user interactions collected between 2013 and 2020 and contains the following information: the queries, the clicked documents, and document attributes. The document attributes include the document title, the document URL, and the document type. For this dataset, we filter the click logs by the click count greater than 5, which leads to all unique queries for the query representation learning. Similar to HS dataset, we specifically set aside the data from January 2018 and February 2018, which is not included in the representation learning dataset for session-based search intent classification.

\section{Co-Query Set Extractor}
The Co-Query Set Extractor groups queries into different sets based on their interactions with clicked documents. It analyzes clicked document annotations to identify patterns and similarities among queries. This approach clusters queries that exhibit similar user engagement or seek related information. For instance, if multiple queries lead users to click on the same set of documents (i.e. same document type or document URL pattern), these queries are considered related and grouped together.

\section{Intent Categories and Annotation}
Understanding user intent is crucial for delivering personalized healthcare experiences. This process necessitates a meticulous approach, beginning with the development of a comprehensive query intent taxonomy, followed by the precise annotation of queries according to these defined intents. 

\subsection{Click-through logs}
Click-through logs track which documents users click on after submitting a query. By analyzing these logs, we can identify patterns in user behavior and infer their intent based on the types of documents they click on most frequently. This approach allows us to efficiently label a large volume of data while still incorporating valuable user interaction signals.

\subsection{HS query intent annotation}
Building an effective healthcare search model hinges on high-quality training data that accurately reflects user intent. However, manually labeling vast amounts of data with accurate intent categories is a significant challenge. To address this, we leverage a semi-automated approach that analyzes user interaction data and the website's organized document structure. Many websites, including healthcare portals, follow a hierarchical structure for organizing content. For example, all pages related to healthcare providers might be found under a URL path like "/provider/", while facility information might reside under "/facilities/". Similarly, account management features could be grouped under a secure section like "/secure/". By understanding this underlying structure, we can define patterns within URLs. First, we take all the URLs in the corpus and define URL patterns (e.g., r"message|email|compose|inbox") for each intent category (e.g., "Communication"). These patterns act as indicators of user intent based on the topmost clicked URL for the given query. For instance, users searching for "note to doctor" consistently clicking URLs with "email" suggests "Communication" intent. By matching top-clicked URLs with predefined URL patterns, we assign intent labels to queries. Search queries can sometimes be open-ended and have several underlying goals (intents) that users might have in mind. This can be observed from the top-clicked results for a particular query. These results might cover various aspects of the topic, indicating different user intents. This allows us to assign ambiguous queries with multiple labels, reflecting the diverse user goals behind a single search phrase.

\subsection{TripClick query intent annotation}
In the TripClick dataset, detailed document annotations provide insights into search intent. We analyze the click patterns of queries on different document types and use these interaction data to infer search intent. This involves collecting statistical data on query clicks for each document type. These clicks are then weighted to reflect their importance and associated with relevant document types as intents in the test dataset.

\section{Detailed Derivation of the Multiset Loss Function}

The multiset loss function is proposed to mitigate the impact of noise present in user co-click log data. This noise primarily arises from randomly co-clicked query pairs, which can obscure the true intent behind user searches. To address this, the multiset loss function employs a clustering-based approach, contrasting batches of search queries within the same cluster against those in different clusters. This method allows for a more meaningful grouping of queries based on user click patterns by utilizing the clicked document annotations, enhancing the accuracy of the representation learning process.

The mathematical formulation of the multiset loss, denoted as \(l_{\text{multiset}} \), is presented in Equation \ref{eq:weighted multiset loss}. Here, \(K\) denotes the number of unique clicked document sets. For each label \(i\), \(N_i\) is the number of queries associated with that label, \(\mathbf{E}_{i_j}\) represents the embedding of the \(j^{th}\) query in document set \(i\), and \(\mathbf{C}_i\) is the centroid of embeddings for set \(i\). The loss function computes the ratio of the average intra-cluster cosine similarity to the average inter-cluster cosine similarity, thus encouraging the model to form tightly knit clusters of queries with shared intents while distancing those with differing intents.

The "sets" used in the multiset cosine similarity loss could be document types or document topics. The loss is calculated as the negative log of the ratio of the average cosine similarity of each set to the average cosine similarity of all sets. With this design, when $N$ is the average number of queries in each document set in the dataset, the theoretical computational cost of multiset loss is $O(K^2*N)$. As opposed to pairwise loss, which has a theoretical complexity of $O(K*N^2)$, multiset loss is computationally more efficient, given that \( K \ll N \), in most conditions.

\section{Detailed Derivation of Training and Evaluation Metrics}

\subsection{Perplexity}
Perplexity measures the intent distribution of a given query, its intent distribution can be represented as a series of probabilities \( P(label_1), P(label_2), ..., P(label_n) \), which is calculated from query click weights in search logs. The calculation of Perplexity is based on these probability values and is defined as:
\begin{equation}
\text{Perplexity} = 2^{-\sum_{i=1}^{n} P(label_i) \log_2 P(label_i)}
\end{equation}

\subsection{Precision}
Precision measures the proportion of correctly predicted positive observations to the total predicted positives for each query:
\begin{equation}
    Precision = \frac{TP}{TP + FP}
\end{equation}
where \( TP \) is the number of true positives and \( FP \) is the number of false positives.

\subsection{F1 Score}
The F1 score is the harmonic mean of Precision and Recall, providing a balance between them. It is particularly useful in scenarios where we have an uneven class distribution. The F1 score is given by:
\begin{equation}
    F1 = 2 \times \frac{Precision \times Recall}{Precision + Recall}
\end{equation}

\subsection{Hit Rate@3}
Hit Rate@3, often used in ranking problems, measures the proportion of times the correct label is within the top 3 predictions. It is a recall-based measure at the top of the ranking list and is calculated as follows:
\begin{equation}
    HitRate@3 = \frac{1}{N} \sum_{i=1}^{N} I(y_i \in top3(\hat{Y}_i))
\end{equation}
where \( N \) is the number of samples, \( y_i \) is the true label for the \( i^{th} \) sample, \( \hat{Y}_i \) is the set of top 3 predicted labels, and \( I \) is the indicator function.

\subsection{Normalized Discounted Cumulative Gain at Rank 3 (NDCG@3)}
Normalized Discounted Cumulative Gain at rank 3 (NDCG@3) evaluates ranking quality, considering the position of the correct intent label in the predicted ranking list. Higher ranks receive more weight. NDCG is calculated as:
\begin{equation}
    NDCG@3 = \frac{1}{N} \sum_{i=1}^{N} \frac{DCG@3_i}{IDCG@3_i}
\end{equation}
where \( DCG@3_i \) is computed from the predicted probability vector \( \mathbf{y}_i \) against the true label vector, representing a weighted vector of correct classification. \( IDCG@3_i \) is the ideal ranking gain, the highest possible \( DCG@3 \) given the set of intent labels. The DCG computation considers the relevance score derived from ground truth \( \mathbf{y}_i \) and the position of the label in the predicted vector \( \mathbf{\hat{y}}_i \).

\end{document}